\documentclass[apj,twocolumn]{emulateapj}
\newcommand{\msolar}{$\rm M_{\odot}$} 
\newcommand{\lsolar}{$\rm L_{\odot}$}
\newcommand{\rsolar}{$\rm R_{\odot}$}
\newcommand{\teff}{$T_{\rm eff}$}
\newcommand{\kms}{km s$^{-1}$}
\renewcommand\email\texttt

\usepackage{amsmath}
\usepackage{txfonts}
\usepackage{amssymb}
\usepackage{graphicx}
\usepackage{color}
\usepackage{longtable}

\begin{document}

\shorttitle{Radial Velocity Curves of Ellipsoidal Red Giant Binaries in the LMC}
\shortauthors{Nie \& Wood}
\title{Radial Velocity Curves of Ellipsoidal Red Giant Binaries
in the Large Magellanic Cloud}
\author{
J. D. Nie\altaffilmark{1,2},
P. R. Wood\altaffilmark{2}}
\altaffiltext{1}{Key Laboratory of Optical Astronomy, National Astronomical Observatories, 
Chinese Academy of Sciences, Beijing 100012, China;\email{jdnie@bao.ac.cn}}
\altaffiltext{2}{Research School of Astronomy and Astrophysics, Australian National
University, Cotter Road, Weston Creek, ACT 2611, Australia;\email{peter.wood@anu.edu.au}}

\begin{abstract}
Ellipsoidal red giant binaries are close binary systems where an
unseen, relatively close companion distorts the red giant, leading to
light variations as the red giant moves around its orbit. These
binaries are likely to be the immediate evolutionary precursors of close
binary planetary nebula and post-asymptotic giant branch and post-red
giant branch stars. Due to the
MACHO and OGLE photometric monitoring projects, the light variability
nature of these ellipsoidal variables has been well studied. However,
due to the lack of radial velocity curves, the nature of their masses,
separations, and other orbital details has so far remained largely
unknown. In order to improve this situation, we have carried out
spectral monitoring observations of a large sample of 80 ellipsoidal
variables in the Large Magellanic Cloud and we have derived radial velocity curves. At
least 12 radial velocity points with good quality were obtained for
most of the ellipsoidal variables. The radial velocity data are provided
with this paper. Combining the photometric and radial velocity data, 
we present some statistical results related to the binary properties 
of these ellipsoidal variables. 
\end{abstract}

\section{Introduction}
The variable red giants in the Large Magellanic Cloud (LMC) fall on six or more distinct
sequences in a period--luminosity (PL) diagram, i.e., 
$K$--log$P$ diagram \citep{1999IAUS..191..151W,2004MNRAS.353..705I,
2007AcA....57..201S,2008AJ....136.1242F}. One of these sequences,
sequence E, consists of binary systems that are mainly red giant
ellipsoidal binaries \citep{1999IAUS..191..151W, 2004AcA....54..347S}.
In these ellipsoidal binary systems, the red giant is the primary and
it substantially fills its Roche lobe. The secondary, which is
usually an unevolved main-sequence star, is unseen observationally in
most cases. Due to the substantial filling of the Roche lobe, the red
giant is distorted. Rotation of the distorted shape of the red giant
as the binary progresses around its orbit causes a change in the
apparent light seen by a distant observer. This leads to the
characteristic light and velocity curves of ellipsoidal variables
that have two cycles of light variation in one orbital period but
only one cycle of radial velocity variation
\citep[e.g.][]{2010MNRAS.405.1770N}

The sequence E stars are low-mass
stars ($m <$ 1.85~\msolar) or intermediate-mass stars 
(1.85~\msolar $ < m <$ 7.0~\msolar). They can lie on either 
the red giant branch (RGB) or the asymptotic giant branch (AGB) and they are therefore in an evolutionary phase where 
the stellar radius is increasing.  It is when the radius of the expanding 
red giant becomes a significant fraction of the binary separation that the 
ellipsoidal variability becomes detectable with current surveys such as MACHO 
and OGLE.  The orbital periods of sequence E stars in the LMC lie in the range
$\sim$30--1000 days and the light amplitude is usually $<$0.3 mag in
the MACHO red band $M_R$. Statistically, the sequence E stars make up
approximately 0.5--2\% of the RGB and AGB stars in the LMC. About 7\%
of the sequence E stars are eclipsing and about 10\% of them have
unusually shaped light curves which indicate significant eccentricity
of the system orbits \citep{2004AcA....54..347S}.

Due to the MACHO and OGLE projects, the variability nature of sequence
E stars is well studied. However, at the present time, there are few
radial velocity studies of the sequence E stars, due to the
difficulties of long-time spectral monitoring.
\citet{2006MmSAI..77..537A} observed two sequence E stars in the LMC,
and their preliminary results indicated that the red giant is filling
its Roche lobe and transferring mass to the companion since the
derived mass of the red giant is close to that of the red giant
core. \citet{2010MNRAS.405.1770N} carried out spectral observations on
11 sequence E stars \citep[including the two in][]{2006MmSAI..77..537A} to 
derive radial velocity variations. The average full velocity amplitude derived 
for those 11 sequence E samples was 43.4 \kms, consistent with the inference that 
sequence E stars are red giants in binary systems with roughly 
solar-mass components. 

A problem with the current understanding of sequence E stars is that
the eccentricity should be close to zero due to tidal interaction with
the companion, yet about 10\% of them have significant eccentricities
\citep{2004AcA....54..347S}.  Complete orbital solutions for ellipsoidal
red giant binaries with a range of eccentricities could possibly 
show in which part of parameter space eccentricity can be maintained.
As a start to such a study, \citet{2012MNRAS.421.2616N} monitored the radial
velocities of 7 eccentric sequence E stars and derived complete orbital 
solutions.

In general, a knowledge of the complete set of orbital properties
(masses, separations, eccentricities, orientations) for a large sample
of ellipsoidal red giant binaries in the LMC will provide a good
resource for understanding the evolution of these close binary
systems.  These observed parameters can be used to constrain Monte
Carlo simulations of the population of sequence E stars, such as those
in \citet{2012MNRAS.423.2764N}.  These Monte Carlo calculations
provide estimates for the production rate of binary post-AGB
stars, binary planetary nebulae (PNe), binary post-RGB stars, and
luminous white dwarfs relative to the production rate of single-star
post-AGB stars and PNe. The relative numbers of the
various types of objects listed above, now quite well known in the
LMC, can be used to parameterize the binary interaction process, 
especially common envelope evolution, and to calibrate models for the 
formation of cataclysmic variable stars, AM CVn systems, the progenitors 
of SNe Ia, etc. So knowledge of the parameters of a large sample
of the precursor binaries should help us better understand the interaction 
processes and the evolutionary fates of close red giant binaries.

In this study, we present the results of radial velocity monitoring of
a large sample of ellipsoidal variables in the LMC.  
In Section \ref {reduction}, the design of our observing
project and the data reduction is described. In Section
\ref{results}, we present observed properties of binary stars based on the radial
velocity and light variation data.  Some statistical analyses are also provided
and compared with theoretical models.
\section[]{Observations and Data Reduction}\label{reduction}

\subsection[]{Selection of Objects}\label{seln_of_objects}
We initially selected 86 sequence E candidates from those given in
\citet{2004AcA....54..347S}.  Only objects with 
$I < 16.5$ mag were considered since good radial velocities could not 
be obtained in a reasonable time (20 minute exposure) for fainter objects.  
Also, the data quality of fainter stars is relatively poor in the MACHO 
and OGLE databases, so their light curves have large dispersions and 
this is not good for fitting orbital solution.

\begin{figure}
\epsscale{1.0}
\plotone{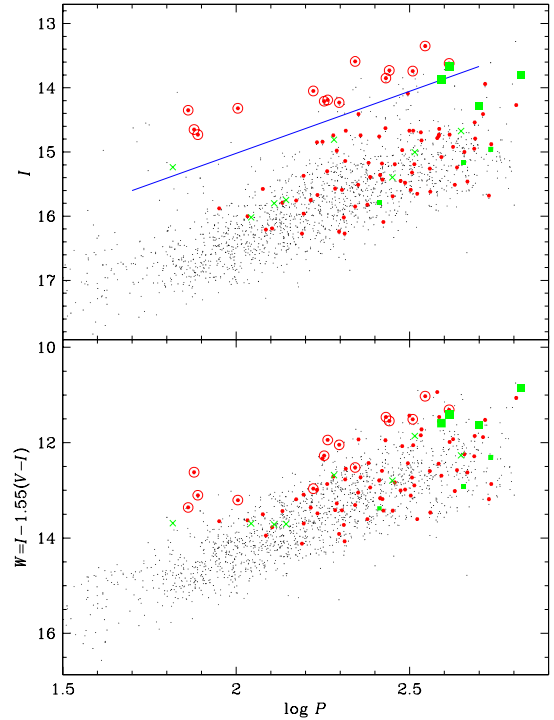}
\caption{Location of our 86 ellipsoidal variable candidates (red dots) 
on the $I$--$\log P$ and $W$--$\log P$ planes.  The small black dots show the full
sample of ellipsoidal variables in \citet{2004AcA....54..347S}. The
blue line in the top panel is a line parallel to the main part of
sequence E.  Stars above this line are brighter than normal for their
period and are expected to be intermediate-mass stars.  Objects in our
sample that lie above the line are shown in both plots as circled
points.  The green squares are ellipsoidal variables studied by 
\citet{2012MNRAS.421.2616N} and the area of the square is proportional 
to the mass of the red giant derived by them. The green crosses are 
ellipsoidal variables studied by \citet{2010MNRAS.405.1770N}. Due to the 
lack of $V$ band data, 3 objects of \citet{2010MNRAS.405.1770N} are absent.}
\label{selection}
\end{figure}

All the sequence E candidates in \citet{2004AcA....54..347S} and our 86 
ellipsoidal variable candidates are plotted in the $I$--$\log P$ and $W$--$\log P$ 
planes in Figure~\ref{selection}, where $W = I - 1.55(V-I)$ is a reddening-free
Weisenheit index \citep{1982ApJ...253..575M}.  It should also be noted
that $P$ is the orbital period, which is twice the period obtained by
Fourier analysis of raw light curves since ellipsoidal variables have
two maxima and two minima of the light curve per orbital
period.  The sequence E variables are often plotted in
PL diagrams using the semi-period. 

Our aim was to select a sample of objects with as wide a set of
parameters as possible.  We therefore selected objects randomly
throughout sequence E, as shown in Figure~\ref{selection}, subject to
the constraint $I < 16.5$. 
Note that the magnitude cut of 16.5 is about 2 mag below
the RGB tip, so our sample includes low-mass sequence E stars,
even down to the masses of globular cluster red giants.
We also made no attempt to select highly eccentric stars so the sample should 
include a range of eccentricities \citep[note that][have already studied 
a sample of 7 eccentric sequence E stars]{2012MNRAS.421.2616N}.  

At a given orbital period, more massive binary
systems will have larger separations so the red giant will need to get
to a higher luminosity (radius) before the Roche lobe filling factor
is large enough to produce detectable ellipsoidal
light variations.  The more massive binaries should therefore
preferentially lie in the high-luminosity side of sequence E in
Figure~\ref{selection}.  The results of \citet{2012MNRAS.421.2616N},
shown in Figure~\ref{selection}, indicate that the higher-mass stars 
do indeed lie on the high-luminosity side of sequence E.
In order to be sure of selecting a good sample of intermediate-mass stars
($M > $$\sim$1.85\,M$_{\odot}$), we preferentially included extra stars 
above the blue line in Figure~\ref{selection}.  Our final sample 
consists of 86 stars.  

With 86 ellipsoidal candidates, plus 11 sequence E stars whose radial velocities
were obtained by \citet{2010MNRAS.405.1770N} (green crosses in
Figure~\ref{selection}) and the 7 eccentric sequence E stars studied 
by \citet[][]{2012MNRAS.421.2616N} (green squares in
Figure~\ref{selection}) , we have a combined sample of approximately 100.
The full sample will give us statistical information about properties such as masses, 
mass ratio, eccentricity, separation, Roche lobe filling factor, and red giant luminosity
for field red giants in the LMC.  Derivation of some of these properties will 
require the use of a binary orbit modeling tool such 
as the Wilson--Devinney code 
\citep{1971ApJ...166..605W, 1979ApJ...234.1054W, 1990ApJ...356..613W,2009ApJ...702..403W}.

\subsection[]{Spectral Observation}

The radial velocity observations were taken using the Wide Field
Spectrograph \citep[WiFeS; ][]{2007Ap&SS.310..255D,2010Ap&SS.327..245D}
mounted on the Australian National University 2.3m telescope
at Siding Spring Observatory. WiFes is an integral field, double-beam,
imaging-slicing spectrograph with a field of view 25$\times$38 square
arcsec, imaged onto 25 slits that are 1 arcsec wide and 38 arcsec
long. It has six gratings, giving high (R=7000) and low (R=3000)
spectral resolutions. For our observations, the gratings B7000
(wavelength coverage of 4184--5580\AA ) and I7000 (wavelength coverage
of 6832--9120\AA) were chosen for the blue and red CCD, respectively.
These two gratings give a two-pixel resolution R=7000, corresponding
to a 45 km s$^{-1}$ velocity resolution.

We carried out 18 weeks of radial velocity monitoring, 
from 2010 September to 2012 March, to cover at least one
orbital period for nearly all of the 86 objects. The observations were
approximately evenly distributed throughout the 18 months, roughly one
radial velocity observation per star per month. The exposure time was 
generally set to 300 s for objects with $I$$\sim$13 mag, 600 s for $I$$\sim$14 
mag, and 900 s for $I$$\sim$15 mag, which gives a signal-to-noise ratio (S/N)  of at least 20. For 
fainter objects ($I > $16 mag), the exposure time was increased to 1200 s. We 
also increased the exposure time when observing in bad weather, to guarantee 
a S/N of 20.

For our observations, we chose the ``stellar" mode exposure, in which case
only 12 slits of the spectrograph were used.  For flatfielding the
QI-1 lamp was used and for wavelength calibration a Ne--Ar arc lamp
exposure was taken at the beginning of the night.  For velocity
derivation, the radial velocity standard star HR9014 was observed.  In
addition, the white dwarf star EG131 was observed so that telluric
lines could be used to remove any zero point error in the wavelength
calibration arising from spectrograph drift over the night.  To
achieve high S/N, the exposure times of HR9014 and EG131 were set to
10 s and 900 s, respectively.

\subsection[]{Data Reduction}

\subsubsection[]{Spectrum Reduction}

The WiFes data reduction pipeline \citep{2010Ap&SS.327..245D} was used
for the spectrum reduction.  The pipeline combines the calibration
and science data and provides one-dimensional spectra with most of the cosmic rays
and sky emission lines removed.  Since all our objects are red, most
of their flux is concentrated in the red region of the spectrum.
Therefore, we did not reduce the blue beam spectra because these
spectra are of low S/N.  The red spectra (6832--9120\AA) contain many
prominent telluric absorption lines as well as emission lines of water, OH, and
O$_2$.  The pipeline-reduced spectra sometimes contained some residual telluric
lines and cosmic rays.  Because of this, we checked by eye all the
reduced spectra of the same object and, by comparison, removed
residual cosmic rays and residual sky emission lines manually.

Spectrally, red giant binaries with white dwarf 
or neutron star companions always show characteristic emission lines \citep{1984PASAu...5..369A,
1986syst.book.....K,1997A&A...327..191M,2000A&AS..146..407B}.
These objects are the symbiotic stars.
In passing, we note that none of our ellipsoidal variables show
emission lines in their spectra. This means that none of the
companions are likely to be white dwarfs or neutron stars.

\subsubsection[]{Radial Velocity Calculation}

\begin{enumerate}
\renewcommand{\theenumi}{(\arabic{enumi})}

\item \emph{Relative radial velocity} 
The relative radial velocity and its error were computed using the
IRAF $fxcor$ package. The $fxcor$ package uses a Fourier cross-correlation method
to find the wavelength shift between an object and a template spectrum
in a specified cross-correlation region.  The template was usually a
radial velocity standard star with well-determined radial velocity and
a spectral type similar to that of the program object. In our case,
all the 86 ellipsoidal variables are stars of K--M spectral type, so
we choose as the template HR9014, which is a K5 star with a well-determined 
radial velocity of $-20.4$ km s$^{-1}$.  The cross
correlation was made on the wavelength interval 8400--8750 \AA~which
contains the Ca II triplet lines and is relatively free of telluric
lines.  The heliocentric radial velocity ($\rm{\upsilon_{helio}(obj)}$), 
obtained from $fxcor$ after heliocentric correction and the inclusion of the 
heliocentric radial velocity of HR9014, was saved along with its error. 
The error in $\rm{\upsilon_{helio}(obj)}$ is normally below 4 km
s$^{-1}$.  If the error was larger than 10 km s$^{-1}$ (due to bad
weather and low S/N), then the velocity point was excluded from our
data set.

\item \emph{Zero-point correction}
To check the velocity calibration, all spectra of each program star
were cross correlated with a template consisting of a single spectrum
of the telluric standard star EG131. EG131 is a white dwarf that
radiates almost like a blackbody.  Thus, its spectrum has only
telluric lines, making it an ideal template for the zero-point
correction.  In principle, the relative velocity of the telluric lines
in different spectra should be zero, but, due to the movement of the CCD
system and spectrograph between the taking of the arc spectrum at 
the beginning of the night and the taking of the object spectrum, 
as well as the changes of the atmospheric pressure, there can be
shifts in velocity.  To compute this zero-point velocity shift
$\rm{\upsilon_{zp}(obj)}$, program stars were cross correlated with EG131 in
the wavelength interval of 8120--8370 \AA.  This region is dominated
by telluric lines and it is close to the region of Ca II triplet, so
the zero-point correction in this region should similar to that of Ca
II triplet region. The zero-point velocity correction varied from
about $-$10 to 34 km s$^{-1}$ across the many nights of the observation. 
HR9014 was also cross correlated
with EG131 to obtain its zero point velocity correction ($\rm{\upsilon_{zp}(HR9014)}$).

\item \emph{Absolute radial velocity}
To compute the absolute value for the observed radial velocity $\rm{\upsilon_r}$
including a zero-point correction, we use the formula
\begin{equation}
\label{vhelio}
\rm {\upsilon_r=\upsilon_{helio}(obj)-\upsilon_{zp}(obj)+\upsilon_{zp}(HR9014)}~~.
\end{equation}

\end{enumerate}

The final radial velocity data for 2 objects are given in Table~\ref{two_rv}.
The full table is available online.

\begin{table}
\begin{center}
\caption{Radial velocities $\rm{\upsilon_r}$ and 1$\sigma$ Errors $\sigma_{\rm{\upsilon_r}}$ in \kms.
Stars are identified by their OGLE II R.A. and decl.}
\label{two_rv}
\vspace{0.1cm}
\begin{tabular}{cccccc}
\tableline\tableline
\multicolumn{3}{c}{OGLE 050659.79 -692540.4 }&\multicolumn{3}{c}{OGLE 051256.36 -684937.5}\\
\tableline
 HJD( 2450000+) & $\rm{\upsilon_r}$ & $\sigma_{\rm{\upsilon_r}}$ &  HJD( 2450000+) & $\rm{\upsilon_r}$ & $\sigma_{\rm{\upsilon_r}}$ \\
\tableline 
 5463.06104 & 246.14 &  1.74&   5461.14209 & 237.26& 1.70\\
 5513.99023 & 251.84 &  2.10&   5513.03125 & 247.21& 1.64\\ 
 5563.00781 & 284.43 &  1.48&   5563.07324 & 257.77& 1.54\\
 5590.98877 & 270.53 &  2.71&   5590.06592 & 256.17& 2.27\\ 
 5603.96924 & 257.47 &  2.12&   5647.92578 & 243.94& 1.51\\ 
 5649.00684 & 232.40 &  1.40&   5673.95508 & 229.63& 2.68\\ 
 5678.97363 & 264.23 &  4.53&   5693.92676 & 230.36& 1.53\\ 
 5679.01221 & 250.28 &  4.04&   5765.23242 & 231.89& 2.03\\
 5764.26514 & 246.01 &  3.65&   5846.08301 & 242.70& 1.92\\
 5783.19287 & 239.24 &  1.88&   5867.01807 & 252.29& 1.39\\ 
 5848.15039 & 273.73 &  7.73&   5907.19336 & 257.03& 1.75\\
 5868.03613 & 282.79 &  2.54&   5944.04883 & 255.03& 1.35\\
 5906.11133 & 270.63 &  1.82&   5991.98584 & 243.43& 1.50\\  
 5944.17676 & 242.22 &  3.00&   -          & -     & -\\ 
 \tableline
 \end{tabular}
 \end{center}
 \end{table}

\section[]{Results}\label{results}

\subsection[]{Properties of the Observed Ellipsoidal Variables}

Before presenting the properties of the observed ellipsoidal variables, 
we need to remove candidates that are not clearly ellipsoidal variables with
the help of the observed radial velocity.  
Among our 86 objects,  we found that 80 of them are real ellipsoidal 
variables, showing two light maxima and two minima but only one velocity 
maximum and minimum in one orbital period. The remaining 6 candidates do 
not clearly satisfy this requirement due to their low-velocity amplitude 
relative to the noise so they were removed.  These 6 objects could be 
ellipsoidal variables whose orbital plane lies close to the plane of the 
sky or they could be sequence D stars as these stars lie close to the 
sequence E stars in  the $I$--$\log P$ and $W$--$\log P$ planes and they 
have small velocity amplitudes \citep{2009MNRAS.399.2063N}.
For the remainder of this paper, we  focus on the 80 real ellipsoidal 
variables while we discuss the 6 rejected objects in the Appendix.

Table \ref{sample} shows various properties of the 80 ellipsoidal
variables.  For all objects, the period given has been re-derived
because the original value from \citet{2004AcA....54..347S} is not
accurate enough.  The MACHO and OGLE II light curve data were taken
more than 10 years before our radial velocity data, which was obtained between
years 2010 and 2012.  Because of the long time span between the two sets of
data, a small error in the period can cause a significant error in the
phase of the light curve projected forward by more than 10 years. To
obtain a more reliable period, when OGLE III data existed, 
we combined OGLE III and OGLE II light
curves (covering the interval 1998--2009) and used the phase
dispersion minimization (PDM) method \citep{1978ApJ...224..953S} to
calculate the period from the combined light curve.  If the OGLE III
data was not available for an object, we combined the OGLE II and
MACHO data, covering the interval 1992--2000. 

The effective temperature was calculated by converting $(I-K)_0$ to
$T_{\rm eff}$ using spline fits to the data in
\citet{2000AJ....119.1424H, 2000AJ....119.1448H}. The reason we used
$(I-K)_0$ as the color index is because most of our sequence E
stars are K- or early-M-type stars, and for these red giants $I-K$ varies
much more with $T_{\rm eff}$ than $J-K$ so that photometric
errors are less important for $I-K$.  The mean $I$ magnitude was
derived from the OGLE photometry, while the $K$ magnitude was obtained
from the Two Micron All Sky Survey (2MASS) catalog
\citep{2003yCat.2246....0C} .  To remove reddening, we adopted
$E(B-V)$=0.08 \citep{2006ApJ...642..834K} along with
$E(V-I)=1.38~\times E(B-V)$ \citep{1998ApJ...500..525S}, and
$E(V-K)=2.744~\times E(B-V)$ and $A(K)=0.35~\times E(B-V)$
\citep{1985ApJ...288..618R}.  The bolometric correction BC$_K$ was
calculated from $(I-K)_0$ using the data in
\citet{2000AJ....119.1424H, 2000AJ....119.1448H}  
and the bolometric luminosity was calculated from $K_{\rm 0}$
and BC$_K$, with a distant modulus of the LMC 18.54 \citep{2006ApJ...642..834K}.

Finally, in Table~\ref{sample}, we indicate whether a star is below (a ``1'' in
the last column) or above (a ``2'' in the last column) the higher-mass line
in Figure~\ref{selection}.  The last column in Table~\ref{sample} also
provides notes on unusual variability characteristics as given in
\citet{2004AcA....54..347S}.  Stars for which there is OGLE III data
available are also indicated.  The data in Table \ref{sample} are available online.

\tabletypesize{\scriptsize}
\tabcolsep=0.20cm
 \begin{deluxetable*}{lllllllllccccl}
 \tablecaption{Properties of the observed ellipsoidal variables\label{sample}}
 \vspace{-0.38cm}
 \tablewidth{0pt}
 \tablehead{
\colhead{No}&      \colhead{Object}&      \colhead{$P$}&
\colhead{$I$}&     \colhead{$V$}&         \colhead{$M_B$}&
\colhead{$M_R$}&   \colhead{$K$}&          \colhead{$\Delta{I}$}& \colhead{$\Delta{\rm RV}$}&
\colhead{$L$}&     \colhead{\teff}&       
\colhead{$R$}&     \colhead{Remark}
\\
\colhead{}&        \colhead{(OGLE II Name)}& \colhead{(day)}&
\colhead{(mag)}&   \colhead{(mag)}&          \colhead{(mag)}&
\colhead{(mag)}&   \colhead{(mag)}&          \colhead{(mag)}&          \colhead{(\kms)}&
\colhead{(\lsolar)}&  \colhead{(K)}&       
\colhead{(\rsolar)}&  \colhead{}
}
\vspace{-0.38cm}
\startdata
1 &050107.08--692036.9&  166.8& 14.05& 14.75&      -&      -& 13.14 &0.110 &  70  & 3278 & 5773  &58 & 2,ogle3\\
2 &050222.40--691733.6&  107.8& 16.00& 17.53&  16.85&  15.87& 13.90 &0.070 &  32  & 581  & 3990  &51 & 1\\
3 &050254.15--692013.8&  171.5& 15.67& 17.08&  -    &  -    & 13.82 &0.025 &  35  & 736  & 4234  &51 & 1\\
4 &050258.71--684406.6&  433.8& 15.24& 16.96&  16.85&  15.65& 13.18 &0.040 &  38  & 1149 & 4020  &71 & 1\\
5 &050334.97--685920.5&  101.1& 14.32& 15.04&  14.85&  14.40& 13.36 &0.200 &  110 & 2526 & 5666  &53 & 2\\
6 &050350.55--691430.2&  224.7& 15.51& 17.10&  16.85&  15.78& 13.46 &0.060 &  24  & 896  & 4030  &62 & 1\\
7 &050353.41--690230.8&  323.6& 14.67& 16.36&  16.28&  15.13& 12.65 &0.060 &  7   & 1915 & 4049  &90 & 1,ogle3\\
8 &050438.97--693115.3&  383.2& 14.75& 16.24&      -&      -& 12.99 &0.070 &  14  & 1673 & 4321  &74 & 1\\
9 &050454.49--690401.2&  220.2& 13.59& 14.28&      -&  13.72& 12.79 &0.070 &  120 & 5175 & 6018  &67 & 2,ogle3\\
10&050504.70--683340.3&  238.9& 15.82& 17.25&  16.95&  16.00& 14.05 &0.040 &  40  & 631  & 4327  &45 & 1\\
11&050512.19--693543.5&  204.3& 16.02& 17.50&  17.30&  16.33& 13.92 &0.050 &  40  & 570  & 3988  &51 & 1\\
12&050554.57--683428.5&  270.1& 13.85& 15.39&  15.10&  14.07& 11.85 &0.035 &  26  & 4028 & 4061  &129& 2\\
13&050558.70--682208.9&  264.0& 15.43& 16.88&      -&  -    & 13.48 &0.020 &  22  & 938  & 4128  &60 & 1\\
14&050604.99--681654.9&  201.7& 15.59& 16.99&  17.01&  15.98& 13.69 &0.030 &  30  & 802  & 4174  &55 & 1\\
15&050610.03--683153.0&  318.5& 15.59& 17.19&      -&      -& 13.71 &0.030 &  35  & 798  & 4197  &54 & 1,e\\
16&050651.36--695245.4&  220.2& 15.85& 17.49&  17.30&  16.13& 13.96 &0.045 &  30  & 630  & 4192  &48 & 1\\
17&050659.79--692540.4&  156.4& 15.37& 16.84&  16.74&  15.75& 13.54 &0.070 &  50  & 963  & 4254  &58 & 1\\
18&050709.66--683824.8&  206.1& 15.14& 16.67&  16.65&  15.60& 13.19 &0.060 &  37  & 1223 & 4121  &69 & 1\\
19&050720.82--683355.2&  521.4& 13.94& 15.50&  15.53&  14.42& 11.95 &0.020 &  30  & 3699 & 4072  &123& 2\\
20&050758.17--685856.3&  198.1& 14.23& 15.64&  15.50&  14.56& 12.37 &0.050 &  45  & 2744 & 4210  &99 & 1\\
21&050800.52--685800.8&  276.2& 13.73& 15.14&  15.02&  14.06& 11.87 &0.050 &  42  & 4337 & 4199  &126& 2\\
22&050843.38--692815.1&  121.8& 16.21& 17.67&  17.30&  16.32& 14.23 &0.040 &  23  & 464  & 4098  &43 & 1\\
23&050900.02--690427.2&  156.5& 15.96& 17.42&  17.25&  16.23& 14.14 &0.025 &  16  & 561  & 4273  &44 & 1\\
24&050948.63--690157.4&  75.60& 14.65& 15.96&  15.80&  14.85& 13.06 &0.200 &  26  & 1779 & 4529  &69 & 2\\
25&051009.20--690020.0&  321.3& 15.32& 16.99&  16.79&  15.63& 13.27 &0.027 &  20  & 1067 & 4024  &68 & 1,e\\
26&051050.92--692228.0&  326.2& 15.19& 16.67&  16.40&  15.47& 13.39 &0.035 &  14  & 1128 & 4284  &62 & 1,ogle3\\
27&051101.04--691425.1&  282.5& 15.69& 17.15&  16.95&  15.97& 13.78 &0.030 &  20  & 732  & 4173  &52 & 1,e\\
28&051200.23--690838.4&  641.3& 14.27& 16.34&  16.13&  14.80& 11.89 &0.030 &  14  & 3131 & 3737  &135& 1,e+sr,ogle3\\
29&051205.44--684559.9&  390.5& 15.08& 16.62&  16.57&  15.54& 12.90 &0.060 &  27  & 1382 & 3912  &82 & 1,ogle3\\
30&051220.63--684957.8&  486.5& 14.54& 16.27&  16.10&  14.92& 12.45 &0.070 &  19  & 2202 & 3983  &99 & 1,e+sr,ogle3\\
31&051256.36--684937.5&  341.9& 14.82& 16.82&  16.60&  15.28& 12.50 &0.130 &  29  & 1851 & 3784  &101& 1,ogle3\\
32&051345.17--692212.1&  197.9& 16.24& 17.74&  17.50&  16.38& 14.19 &0.040 &  42  & 459  & 4038  &44 & 1\\
33&051347.73--693049.7&  306.6& 15.48& 17.09&  -    &  -    & 13.52 &0.040 &  20  & 899  & 4114  &60 & 1,e\\
34&051515.95--685958.1&  322.7& 13.74& 15.18&  15.02&  13.98& 11.91 &0.030 &  24  & 4269 & 4229  &123& 2,e\\
35&051620.47--690755.3&  240.7& 15.17& 16.94&  16.85&  15.58& 13.05 &0.150 &  40  & 1249 & 3963  &76 & 1,ogle3\\
36&051621.07--692929.6&  179.3& 14.21& 15.46&  15.17&  14.36& 12.43 &0.030 &  17  & 2748 & 4294  &96 & 2\\
37&051653.08--690651.2&  269.2& 14.63& 16.36&  16.35&  14.93& 12.32 &0.080 &  29  & 2193 & 3792  &110& 1\\
38&051738.19--694848.4&  297.7& 15.45& 17.03&  16.80&  15.73& 13.31 &0.050 &  29  & 975  & 3943  &68 & 1,e\\
39&051746.55--691750.2&  225.2& 14.41& 16.01&      -&      -& 12.33 &0.050 &  15  & 2478 & 3985  &105& 1\\
40&051818.84--690751.3&  463.9& 15.46& 17.29&      -&      -& 13.03 &0.050 &  26  & 1076 & 3710  &80 & 1,e\\
41&051845.02--691610.5&  320.0& 15.65& 16.97&      -&      -& 13.90 &0.040 &  55  & 734  & 4349  &48 & 1\\
42&052012.26--694417.5&  177.4& 14.84& 16.46&      -&  15.23& 12.75 &0.040 &  18  & 1675 & 3983  &87 & 1\\
43&052029.90--695934.1&  171.0& 14.85& 16.05&  15.88&  15.00& 13.23 &0.045 &  50  & 1489 & 4492  &64 & 1\\
44&052032.29--694224.2&  72.70& 14.35& 14.99&  14.73&  14.31& 13.52 &0.060 &  60  & 2569 & 5973  &48 & 2,e\\
45&052048.62--704423.5&  488.5& 14.79&    -&      -&       -& 12.95 &0.030 &  5   & 1641 & 4231  &76 & 1\\
46&052115.05--693155.1&  194.9& 14.98& 16.08&  15.78&  15.03& 13.38 &0.020 &  49  & 1318 & 4525  &60 & 1\\
47&052117.49--693124.8&  243.3& 15.39& 16.97&  16.60&  15.55& 13.21 &0.060 &  38  & 1044 & 3908  &71 & 1\\
48&052119.56--710022.1&  300.9& 14.97& 16.84&  16.65&  15.40& 12.79 &0.010 &  23  & 1531 & 3907  &86 & 1,ogle3\\
49&052203.16--704507.8&  362.5& 15.62& 17.01&  16.72&  15.82& 14.04 &0.035 &  14  & 734  & 4564  &44 & 1\\
50&052228.85--694313.7&  231.0& 14.74& 16.04&  15.77&  14.85& 12.99 &0.030 &  37  & 1684 & 4334  &74 & 1\\
51&052238.43--691715.1&  77.53& 14.73& 15.78&  15.41&  14.74& 13.47 &0.070 &  55  & 1638 & 5046  &53 & 2\\
52&052324.57--692924.2&  312.4& 14.09& 15.16&  15.08&  14.43& 12.39 &0.030 &  40  & 3016 & 4388  &96 & 2\\
53&052422.28--692456.2&  485.4& 14.95& 16.67&  16.65&  15.49& 12.82 &0.060 &  18  & 1535 & 3948  &85 & 1,+sr,ogle3\\
54&052425.52--695135.2&  206.7& 14.67& 16.04&  15.87&  15.03& 12.93 &0.040 &  42  & 1792 & 4345  &75 & 1\\
55&052438.19--700435.9&  192.8& 15.57& 16.96&  16.70&  15.87& 13.79 &0.040 &  25  & 794  & 4314  &51 & 1\\
56&052438.40--700028.8&  410.8& 13.62& 15.11&  14.83&  14.00& 11.71 &0.060 &  50  & 4852 & 4147  &136& 2,e\\
57&052458.88--695107.0&  340.6& 14.67& 16.53&  16.33&  15.20& 12.35 &0.075 &  5   & 2117 & 3790  &108& 1,+sr,ogle3\\
58&052510.82--700123.9&  189.7& 14.74& 16.05&  15.87&  15.05& 12.99 &0.030 &  30  & 1686 & 4328  &74 & 1\\
59&052513.34--693025.2&  259.4& 14.76& 16.03&  15.97&  15.14& 13.12 &0.030 &  45  & 1621 & 4471  &68 & 1\\
60&052542.11--694847.4&  514.6& 14.41& 16.04&  15.90&  14.85& 12.35 &0.030 &  35  & 2461 & 4004  &104& 1,e\\
61&052554.91--694137.4&  182.8& 15.30& 16.87&  16.95&  15.85& 13.23 &0.120 &  40  & 1092 & 4008  &69 & 1\\
62&052703.65--694837.7&  350.4& 13.35& 14.85&  14.70&  13.78& 11.43 &0.100 &  60  & 6230 & 4131  &156& 2\\
63&052833.50--695834.6&  183.2& 14.19& 15.64&  15.65&  14.75& 12.31 &0.060 &  20  & 2861 & 4186  &103& 2\\
64&052928.90--701244.2&  531.5& 15.68& 17.29&      -&      -& 13.71 &0.060 &  25  & 751  & 4105  &55 & 1,e\\
65&052948.84--692318.7&  421.7& 14.92& 16.85&  16.55&  15.35& 12.70 &0.070 &  16  & 1625 & 3873  &90 & 1,e+sr,ogle3\\
66&052954.82--700622.5&  154.7& 16.27& 17.66&  17.48&  16.63& 14.33 &0.050 &  40  & 434  & 4149  &41 & 1\\
67&053141.27--700647.1&  384.1& 14.73& 16.78&      -&      -& 12.36 &0.090 &  24  & 2044 & 3751  &108& 1,e+sr,ogle3\\
68&053156.08--693123.0&  412.2& 14.73& 16.50&      -&      -& 12.57 &0.070 &  22  & 1899 & 3916  &96 & 1,e+sr,ogle3\\
69&053202.44--693209.2&  426.5& 15.51& 17.12&      -&      -& 13.48 &0.030 &  25  & 892  & 4042  &62 & 1,e\\
70&053219.66--695805.0&  89.55& 15.85& 17.32&  16.70&  15.90& 12.14 &0.010 &  18  & 1749 & 3264  &132& 1\\
71&053226.48--700604.7&  454.7& 15.00& 16.78&  16.43&  15.28& 12.72 &0.070 &  40  & 1547 & 3817  &91 & 1,e+sr,ogle3\\
72&053337.07--703111.7&  315.5& 14.67& 16.76&  16.60&  15.28& 12.37 &0.130 &  16  & 2108 & 3799  &107& 1,e,ogle3\\
73&053338.94--694455.2&  126.9& 16.19& 17.75&  17.60&  16.51& 14.23 &0.090 &  32  & 469  & 4128  &43 & 1\\
74&053356.79--701919.6&  205.3& 16.27& 17.69&  17.53&  16.55& 14.35 &0.040 &  40  & 432  & 4169  &40 & 1\\
75&053438.78--695634.1&  384.7& 14.64& 16.69&  16.35&  15.07& 12.28 &0.070 &  19  & 2218 & 3751  &113& 1,ogle3\\
76&053733.21--695026.9&  263.0& 15.18& 16.85&  16.84&  15.63& 13.07 &0.050 &  19  & 1235 & 3970  &75 & 1\\
77&053946.88--704257.8&  265.0& 16.09& 17.81&  17.78&  16.55& 13.93 &0.040 &  41  & 547  & 3929  &51 & 1\\
78&054006.47--702820.4&  361.5& 15.26& 16.98&  16.85&  15.63& 13.08 &0.070 &  29  & 1176 & 3906  &76 & 1,e,ogle3\\
79&054258.34--701609.2&  164.0& 15.75& 17.29&  -    &  -    & 13.79 &0.040 &  20  & 703  & 4114  &53 & 1\\
80&054736.16--705627.2&  135.9& 15.79& 17.31&  17.05&  16.03& 13.96 &0.030 &  35  & 657  & 4253  &48 & 1\\
\enddata
\vspace{-0.38cm}
\tablecomments{Columns 4--8 are mean magnitudes in different bands; Column 9 is the mean amplitude of light 
variability in $I$ band; Column 10 is the full amplitude of radial velocity; Columns 11--13 are the luminosity, 
effective temperature and mean radius derived as described in the text; Column 14 contains remarks: 
number 1 for stars below the line in Figure~\ref{selection} and number 2 for stars above the line, 
lower case letters stand for binary types: ``$e$,"~eccentric, ``$+sr$," a binary
whose red giant shows semi-regular variability as well as ellipsoidal variability, and ``ogle3" denotes an object which has OGLE III light curve data.}
\end{deluxetable*}

\subsection[]{Light Curves and Radial Velocity Curves}\label{rv_curve}

\begin{figure*}
\includegraphics[angle=0,width=0.5\textwidth,height=0.55\hsize]{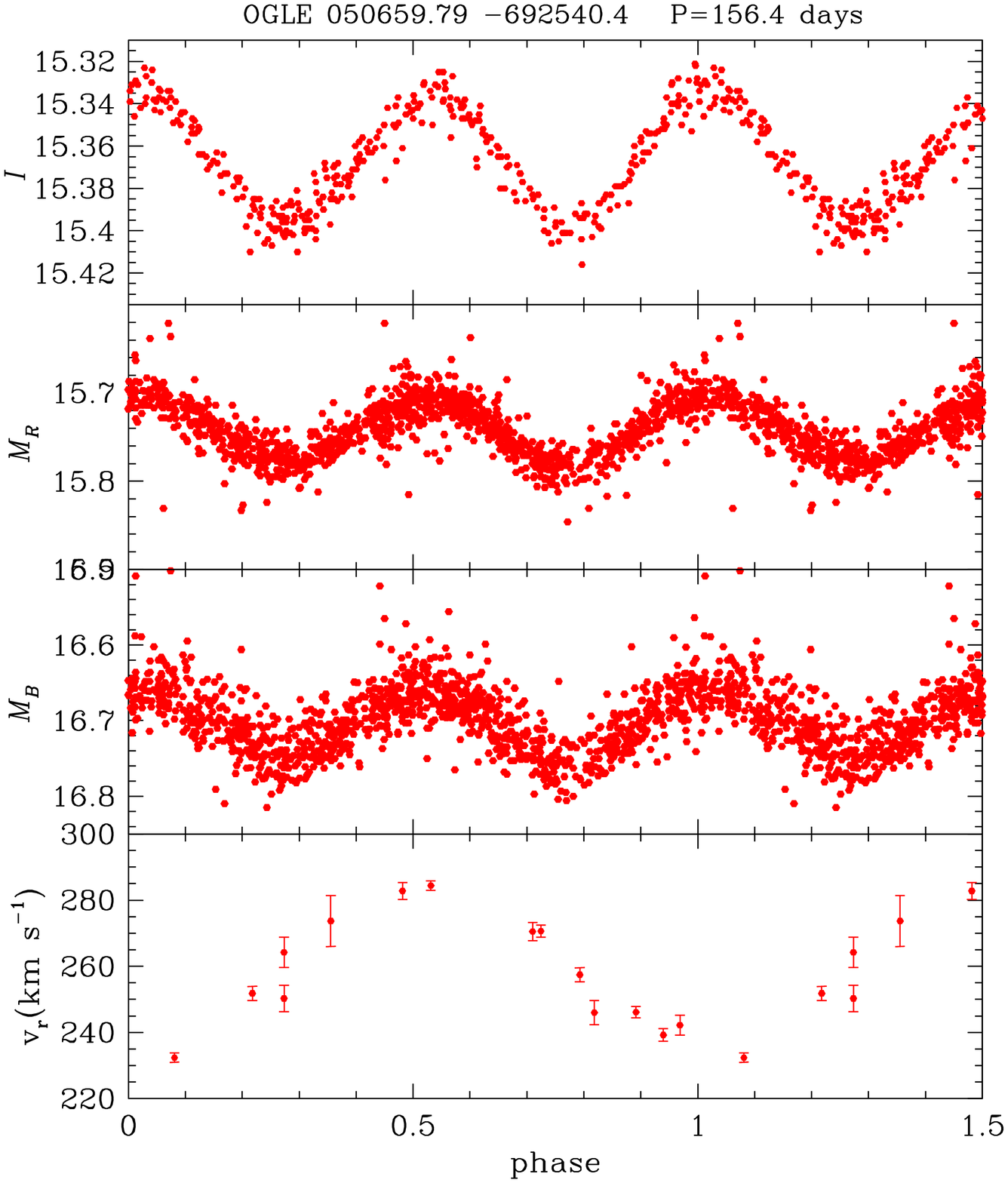}
\includegraphics[angle=0,width=0.5\textwidth,height=0.55\hsize]{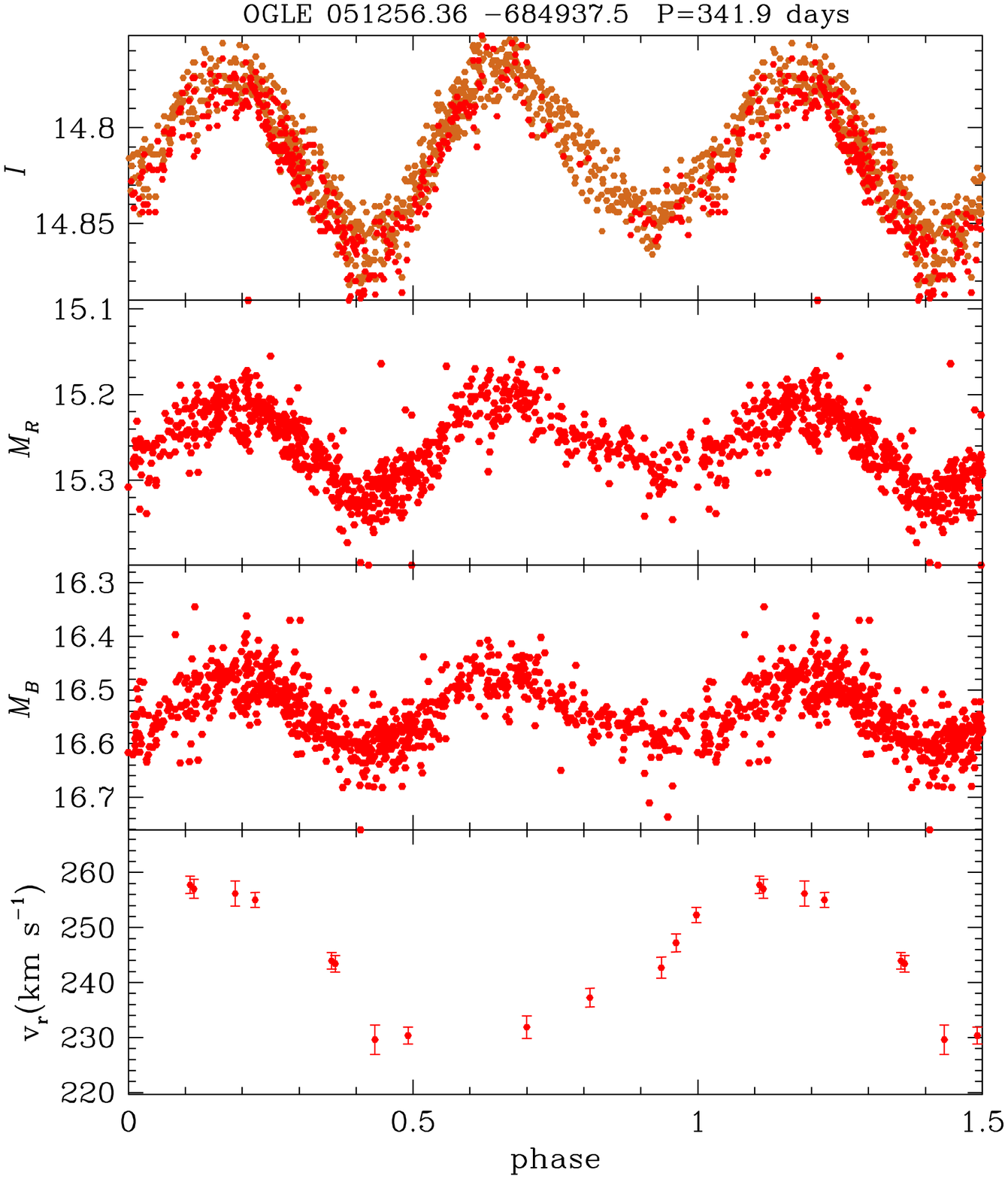}
\includegraphics[angle=0,width=0.5\textwidth,height=0.55\hsize]{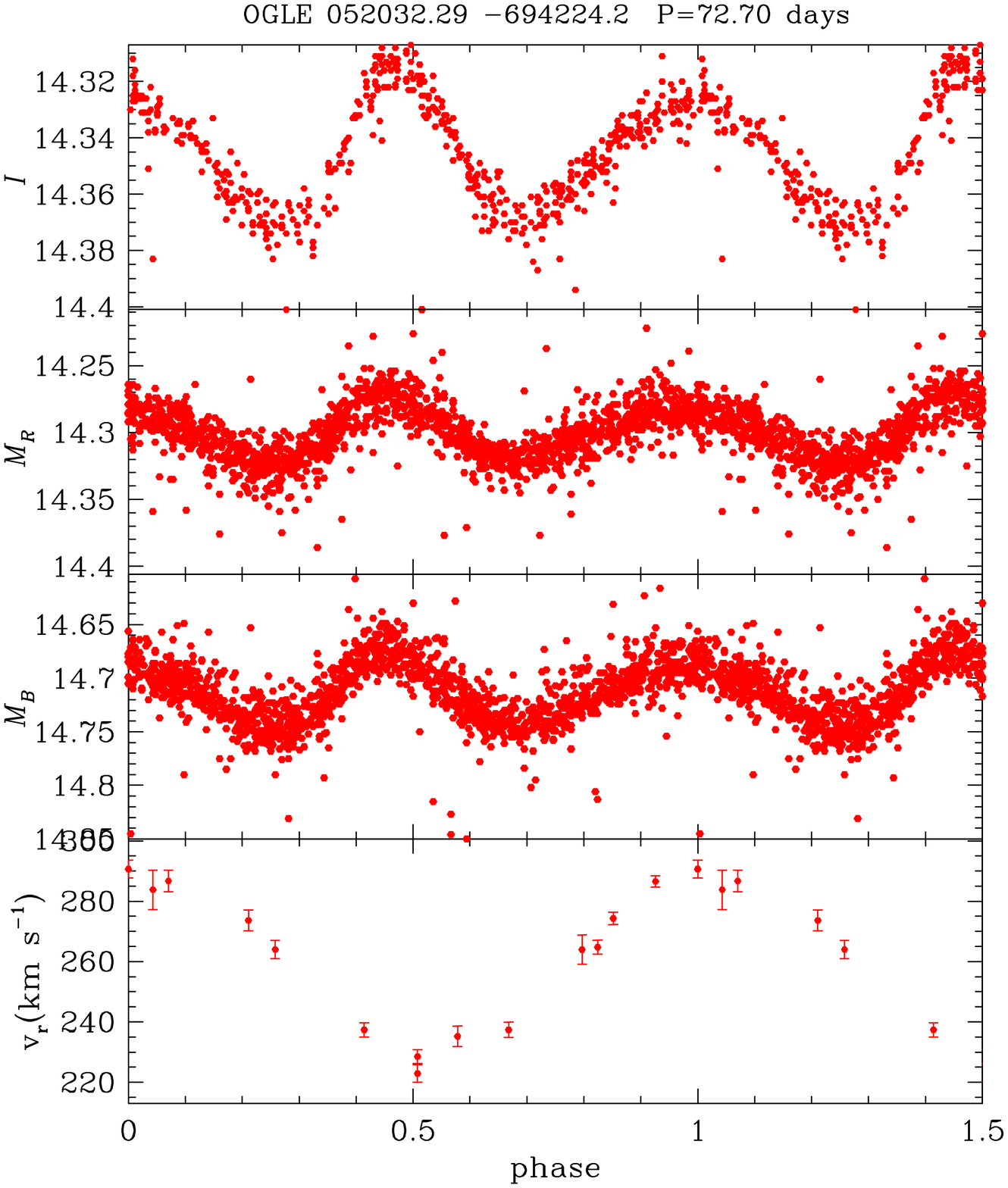}
\includegraphics[angle=0,width=0.5\textwidth,height=0.55\hsize]{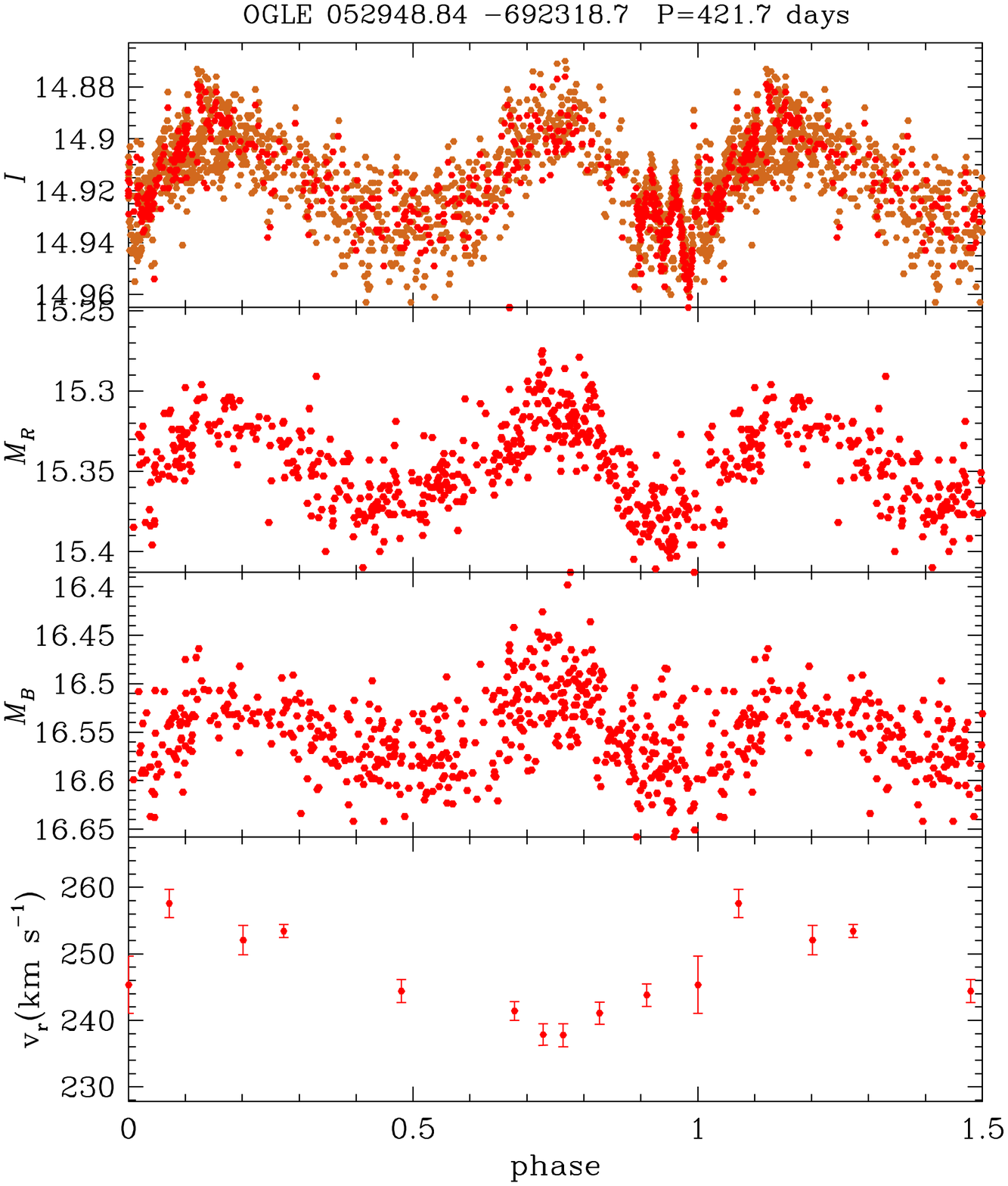}
\caption{Light and radial velocity curves of ellipsoidal variables.
$M_R$ and $M_B$ are the MACHO red and blue magnitudes, respectively.
The top two objects are ellipsoidal variables with circular orbits while 
the bottom two have eccentric orbits. For the $I$ band data, chocolate  points
are from OGLE III and red points are from OGLE II.
\label{fig_lc_rv}}
\end{figure*}

We present samples of light and radial velocity curves for the
ellipsoidal variables in Figure \ref{fig_lc_rv}. Light and radial 
velocity curves for all objects are available as online data.
The time series data for the light curves are from the MACHO, OGLE II, 
and OGLE III databases if they are available. Each radial velocity curve
has at least 12 good quality data points in one orbital period, good
enough for a binary orbital solution.  As confirmed by
\citet{2010MNRAS.405.1770N}, our observed ellipsoidal variables show two light
maxima and two minima in one orbital period, while the velocity shows
only one maximum and one minimum.  Among our 80 objects, most of them
(60/80=75\%) have almost circular orbits, since they have equal light
maxima and they have the same light curve widths for the first and second
maxima (e.g., the top two objects of Figure \ref{fig_lc_rv}).  The other 
systems are eccentric binaries (20/80=25\%) since they show different 
light maxima or light curve widths in one orbital period (e.g., the bottom 
two objects of Figure \ref{fig_lc_rv}).  In addition, there are binary 
systems that contain a semi-regular pulsating red giant (8/80=10\%) 
(e.g., object OGLE 052948.84 
in Figure \ref{fig_lc_rv}), similar to those noted 
by \citet{2012MNRAS.421.2616N}.  For these systems, both pulsation 
theory and binary theory can be used to constrain and compare stellar 
parameters.

\subsection[]{The Velocity Amplitude}

\begin{figure}
\epsscale{1.0}
\plotone{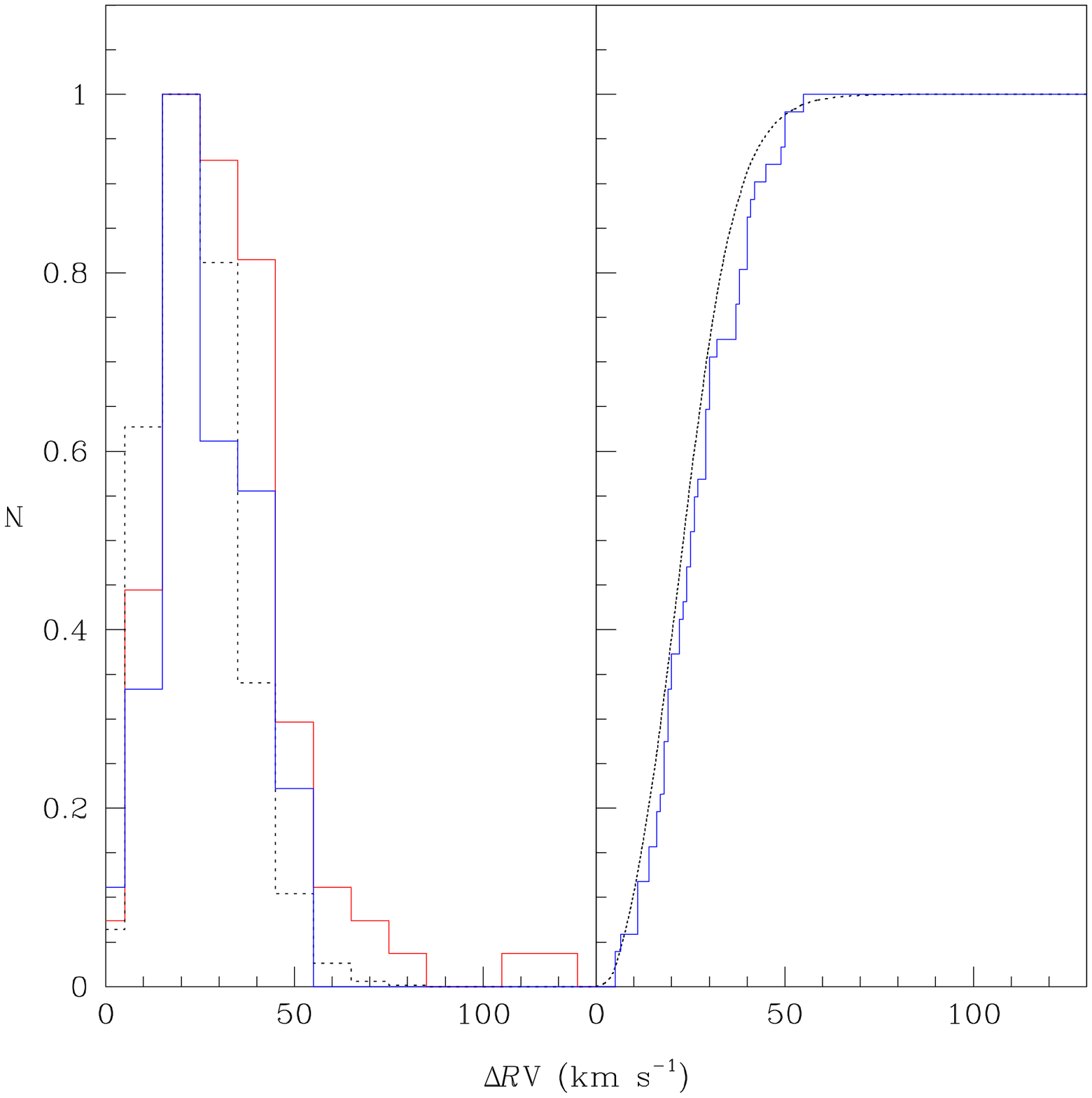}
\caption{
Distribution of the full radial velocity amplitude $\Delta \rm RV$. 
Left panel: the red solid line shows the distribution for the 80 ellipsoidal
variables and the combined sample of \citet{2010MNRAS.405.1770N} and \citet{2012MNRAS.421.2616N}, 
the blue solid line shows the observed distribution for 51 ellipsoidal variables on the top one magnitude of the RGB,
and the black dashed line shows the predicted distribution for  ellipsoidal variables on the top
one magnitude of the RGB by the model of \citet{2012MNRAS.423.2764N} with the calibration of OGLE II data.
Right panel: the predicted cumulative distribution 
for ellipsoidal variables on the top one magnitude of the RGB. Blue is for observations from 51 objects,
and black is for the model of \citet{2012MNRAS.423.2764N} with the calibration of OGLE II data.
Note that all the distribution peaks are normalized to unity.}
\label{fig_rv_amp}
\end{figure}

The distribution of full radial velocity amplitude $\Delta RV$ for the 80 
ellipsoidal variables plus the combined sample from
\citet{2010MNRAS.405.1770N} and \citet{2012MNRAS.421.2616N} is presented in 
the left panel of Figure~\ref{fig_rv_amp} (red). From the figure, it can be seen that the velocity amplitude varies from 
$\sim$5 to $\sim$130 km s$^{-1}$, with a peak near 30 km s$^{-1}$. Note 
that the number of stars with velocity amplitudes larger than 80 km s$^{-1}$ is small. 

Our observed velocity amplitude distribution is also compared to the
model prediction of  \citet{2012MNRAS.423.2764N} (the black dashed 
line in the left panel of Figure 3). The predicted distribution is for 
ellipsoidal red giant binaries on the top one magnitude of the RGB (870--2190\,L$_{\odot}$) 
that have light amplitudes detectable by the OGLE II observations. Our
comparison sub-sample of observations consists of 51 objects for which the 
luminosity lies on the top one magnitude of the RGB (the blue
solid line in the left panel).  Note that  we compare to the 
model of \citet{2012MNRAS.423.2764N} which is calibrated on the OGLE II data 
of \citet{2004AcA....54..347S} because that calibration is better 
than the one based on MACHO data due to the extra sensitivity of OGLE II
observations to small-amplitude light variations.

The cumulative distributions corresponding to the histograms of the
ellipsoidal variables on the top one magnitude of the RGB in 
Figure \ref{fig_rv_amp} are shown in the right panel.  A two-sample 
Kolmogorov--Smirnov test gives a probability of up to 0.20 that the 
observation and model come from the same underlying distribution.  
There is thus a modest probability that the model is consistent with 
both the OGLE II photometry and the velocity amplitudes derived in this study.

\subsection[]{The Mass Function}

\begin{figure}
\epsscale{1.0}
\plotone{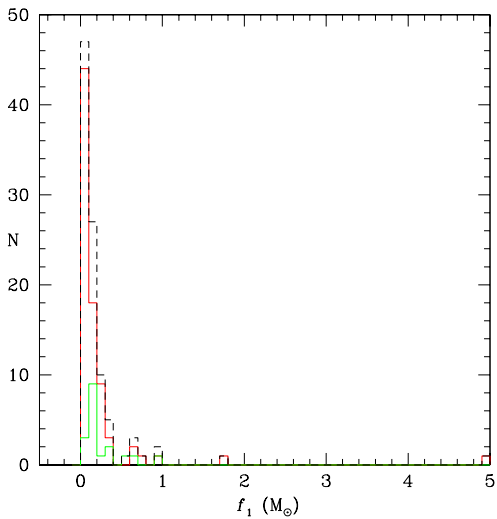}
\caption{Binary mass function distribution for red giant ellipsoidal 
variables in the LMC. The red line denotes our observation, the green 
line denotes the combined observations of \citet{2010MNRAS.405.1770N} 
and \citet{2012MNRAS.421.2616N}, and the black dashed line is the sum 
of the two distributions.}
\label{fig_mass_fun}
\end{figure}

The binary mass function $f_1$ of a binary system where the primary 
star (in our case, the red giant) is observable is
\begin{equation}
\label{mass_function}
f_1=\frac{K_1^3P}{2\pi~G}=\frac{m_2^3\sin^3~i}{(m_1+m_2)^2}~,
\end{equation}
where $K_1$ is the radial velocity semi-amplitude of the primary star, 
$P$ is the orbital period, $m_1$ is the mass of the primary star, and 
$m_2$ is the mass of the secondary star.  Note that $f_1$ is a minimum 
estimate for $m_2$. The distribution of $f_1$ for our sample of red giant ellipsoidal 
variables in the LMC is shown in Figure~\ref{fig_mass_fun}. The figure 
shows a peak at about 0.2~$\rm M_{\odot}$, presumably due to the dominant
low-mass population, but there are a few objects that must be in
more massive binaries where the secondary star is of intermediate-
mass with $m_2 \ga 4~\rm M_{\odot}$.  Red giants of this mass were found 
in ellipsoidal binaries in the LMC by \citet{2012MNRAS.421.2616N}. If we 
assume a random pole orientation for the binary orbit, which implies that 
the mean of $\sin^3 i = \frac{3 \pi}{16}$, and if we also assume that the
mean mass ratio is $q=1$, then for stars in the peak of the distribution 
around $f_1 \sim 0.2~{\rm M_{\odot}}$, the mean mass of the red giant is 
$m_1 \sim$1.35~${\rm M_{\odot}}$.  This is similar to peak in the mass 
distribution of red giants predicted by modeling the star formation 
history of the LMC \citep[e.g.,][]{2012MNRAS.423.2764N}. 

\subsection[]{The PL Diagram}
\begin{figure}
\epsscale{1.0}
\plotone{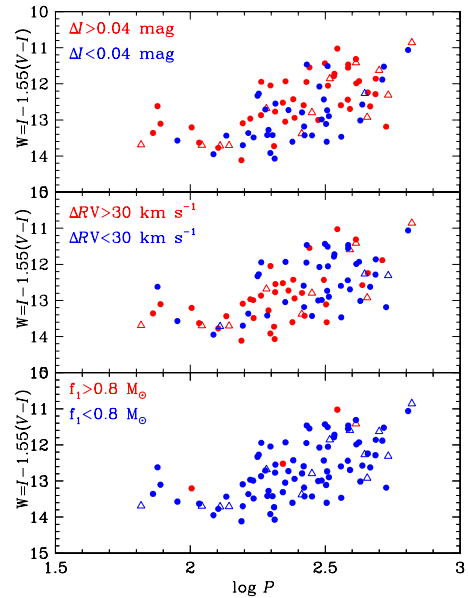}
\caption{PL diagram for the our 80 ellipsoidal variables. The top panel 
shows the position on the PL diagram of objects with different full-amplitude 
$\Delta I$, the middle panel shows the position of objects with 
different full-amplitude $\Delta \rm RV$ and the bottom panel shows the position of 
objects with different mass function $f_1$.  In all panels, filled circles 
denote observations from this work and open triangles denote observations 
from \citet{2010MNRAS.405.1770N} and \citet{2012MNRAS.421.2616N}.}
\label{fig_amp_pl}
\end{figure}

The variations of light amplitude, velocity amplitude and mass function
across the PL diagram are shown in Figure \ref{fig_amp_pl}.  In the top
panel, we can see that the higher-amplitude ellipsoidal variables tend 
to lie on the higher luminosity, shorter period side of sequence E. 
This is just as demonstrated by \citet{2004AcA....54..347S} and is 
presumably because as the red giant in a binary system (with a given 
orbital period) evolves to higher luminosity, it will expand to a 
greater filling fraction of its Roche lobe and hence be more distorted 
and have a large light variation amplitude.

The middle panel of Figure \ref{fig_amp_pl} shows a tendency for higher-
velocity amplitudes to be associated with shorter-period orbits, as might 
be expected.  Of course, low-velocity amplitudes can be seen for shorter-
period orbits if the orbital plane is oriented close to the plane of
the sky: there are a few objects that seem to fall in this category.

The mass function also shows some correlation with position in the
PL diagram (bottom panel of Figure \ref{fig_amp_pl}). The most
massive objects, with $f_1 > 0.8$~\msolar, lie on the upper part 
of sequence E in agreement with the results shown in Figure \ref{selection}.

\subsection[]{Velocity Amplitude versus Light Amplitude}

\begin{figure}
\epsscale{1.0}
\plotone{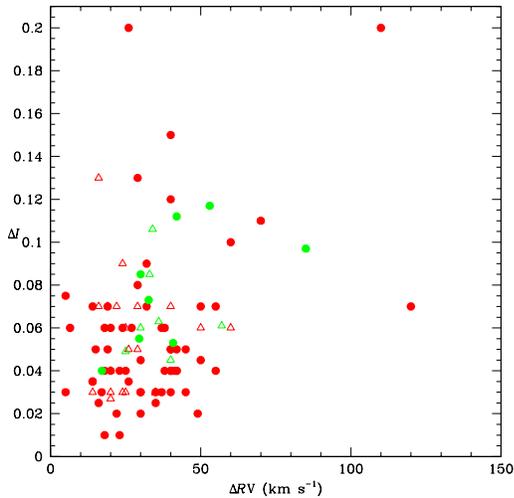}
\caption{Relation between the velocity amplitude and light amplitude. Red: observations
from this work; circles are for circular binaries and triangles are for
eccentric binaries. Green: observations from \citet{2010MNRAS.405.1770N} and 
\citet{2012MNRAS.421.2616N}; circles are for circular binaries and  
triangles are for eccentric binaries. Note that some objects are overwritten 
because they have the same velocity and light curve amplitude.
\label{fig_amprv_ampi}}
\end{figure}

In Figure \ref{fig_amprv_ampi}, we present the light amplitude as a
function of the velocity amplitude. Red symbols are our observation,
and green symbols are observations from \citet {2010MNRAS.405.1770N}
and \citet{2012MNRAS.421.2616N}.  The 20 eccentric ellipsoidal
variables from our observations and the 7 highly eccentric ellipsoidal
variables studied by \citet{2012MNRAS.421.2616N} are marked as open
triangles. As expected, there is generally no correlation of the light
and velocity amplitudes.  The former depends on the Roche lobe filling
factor (and inclination), while the latter depends on the stellar
masses and the orbital separation.  In principal, the orbital
separation could influence the Roche lobe filling factor, but in practice 
the orbital separation effectively determines the luminosity on the giant 
branch where the Roche lobe nearly fills and ellipsoidal variability becomes 
detectable.  This suggests that the orbital separation (velocity amplitude) 
should depend on luminosity for red giant ellipsoidal variables, but the 
light amplitude should not depend on orbital separation (velocity amplitude).

\subsection[]{Velocity Amplitude versus Luminosity}

\begin{figure}
\epsscale{1.0}
\plotone{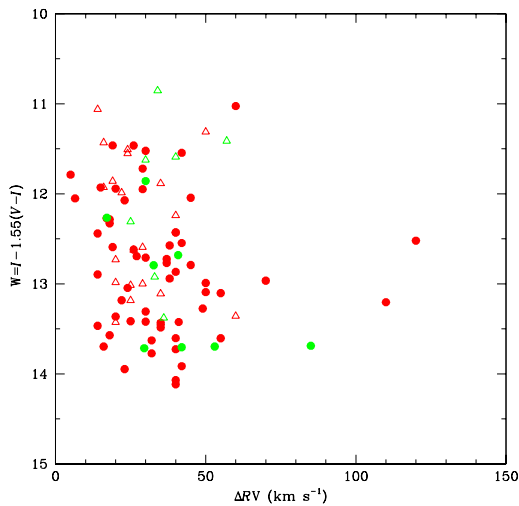}
\caption{Relation between velocity amplitude and luminosity. Symbols are as in
Figure~\ref{fig_amprv_ampi}.  
\label{fig_w_amp}}
\end{figure}

As suggested in the last section, the velocity amplitude should show some
correlation with luminosity. In Figure \ref{fig_w_amp}, we show the 
dependence of luminosity on velocity amplitude.  The binaries with larger 
velocity amplitudes do tend to have lower luminosities although the effect 
is not prominent.  This is as expected, since higher-velocity amplitude 
generally means a smaller separation and thus Roche lobe filling should 
occur at a lower luminosity.

\section[]{Summary and Conclusions}

We have presented radial velocity observations obtained for 80
ellipsoidal red giant binaries in the LMC.  This sample is much larger
than the previous samples of \citet{2010MNRAS.405.1770N} and
\citet{2012MNRAS.421.2616N}.  The mass function of the sample suggests
that the typical mass of the red giants in the ellipsoidal binaries is
$\sim$1.35\,M$_{\odot}$, in agreement with estimates derived from
the star formation history of the LMC.

The main purpose of this paper was to present radial velocity data
and basic observational properties of 80 ellipsoidal
variables in the LMC.  In future work, the radial velocity data will be
combined with MACHO and OGLE photometric light curve data to give complete 
orbital solutions for these binary systems.  This is possible
because the distance, and hence the luminosity, of these LMC objects is well known
\citep[e.g.][]{2012MNRAS.421.2616N}. The results of the complete solutions will 
yield statistical distributions of masses, mass ratios, separations, and eccentricities 
in the period range observed. These distributions of orbital elements for binaries in the LMC 
can be compared with the solar vicinity statistical data given in the 
classic paper of \citet{1991A&A...248..485D} and the recent paper by
\citet{2010ApJS..190....1R}.  It will be interesting to see if the
same distributions exists in samples of binaries with different
metallicity distributions and different star formation histories.

\acknowledgments
We acknowledge constructive comments by the anonymous
referee. J.D. N. thanks Geoff White, Donna Burton, Catherine Farage, 
and the other technical staff at Siding Springs Observatory (SSO) for 
their assistance and support throughout this observing project.
J.D.N. is supported by the National Natural Science 
Foundation of China (NSFC) through grant 11303043 and the 
Young Researcher Grant of National Astronomical Observatories, 
Chinese Academy of Sciences. P.R.W. received partial support for this
project from Australian Research Council Discovery Project grant
DP120103337.

\appendix

\section[]{List of objects rejected}  \label{appendix} 
\begin{figure*}
\includegraphics[angle=0,width=0.34\textwidth,height=0.35\hsize]{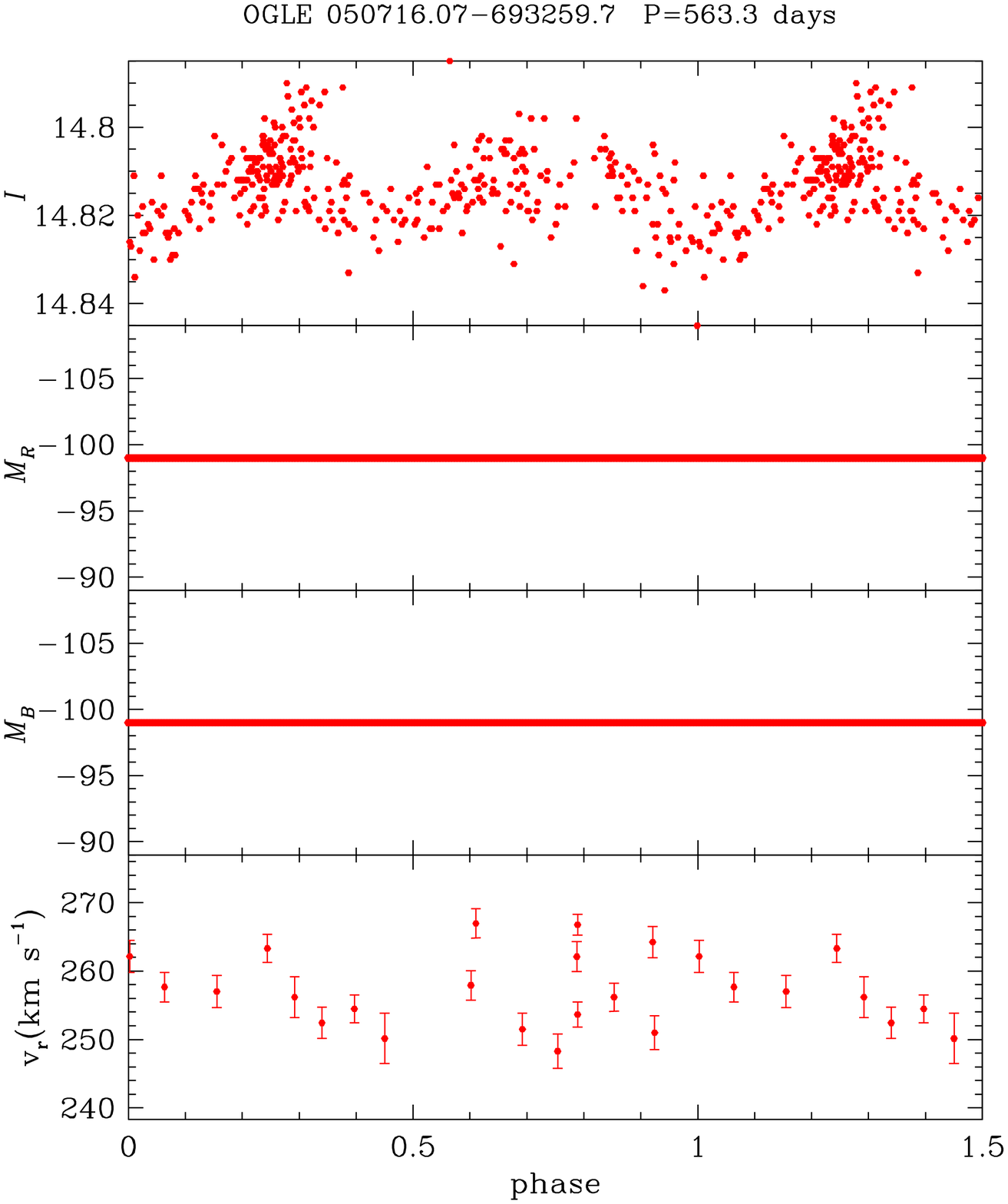}
\includegraphics[angle=0,width=0.34\textwidth,height=0.35\hsize]{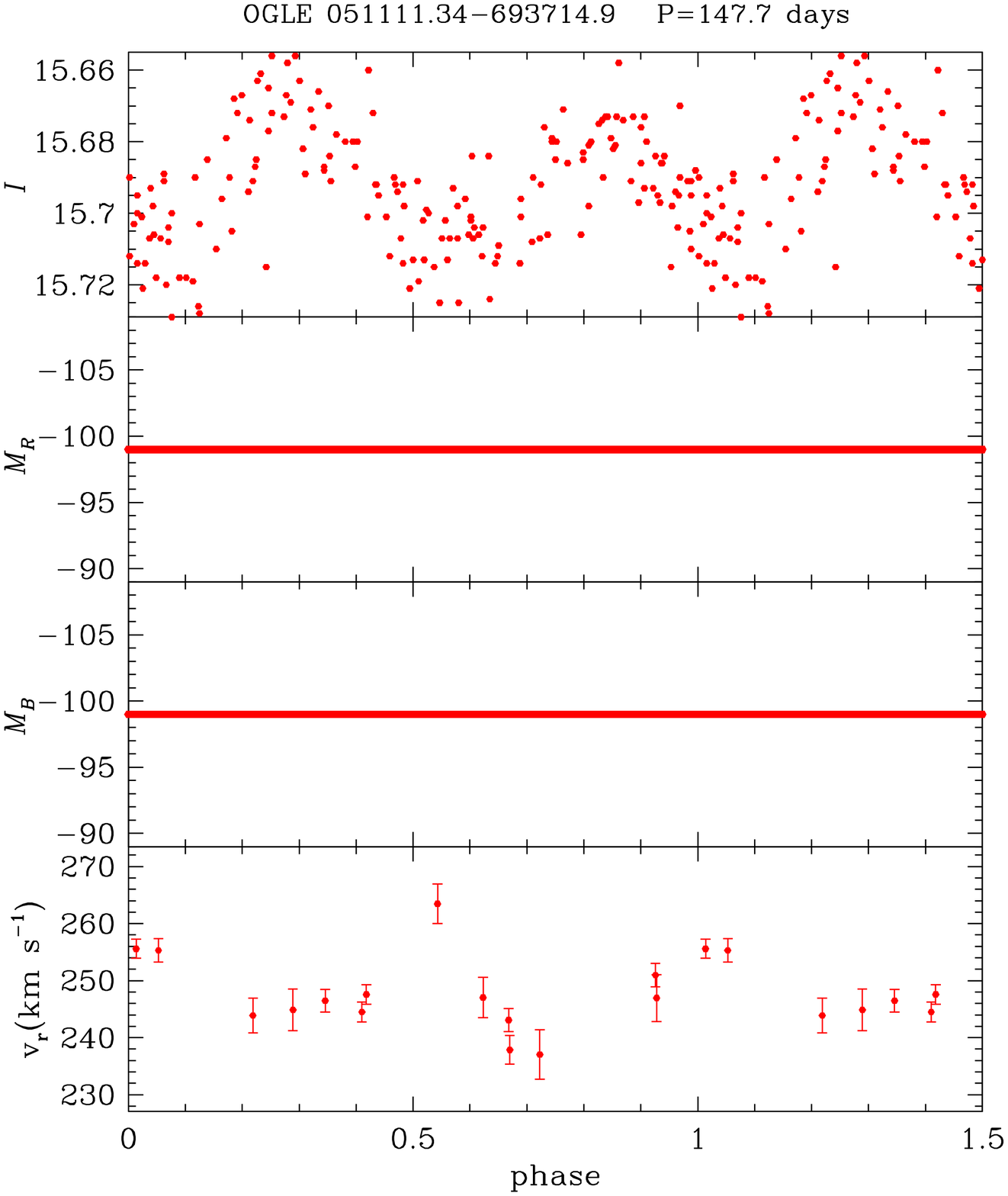}
\includegraphics[angle=0,width=0.34\textwidth,height=0.35\hsize]{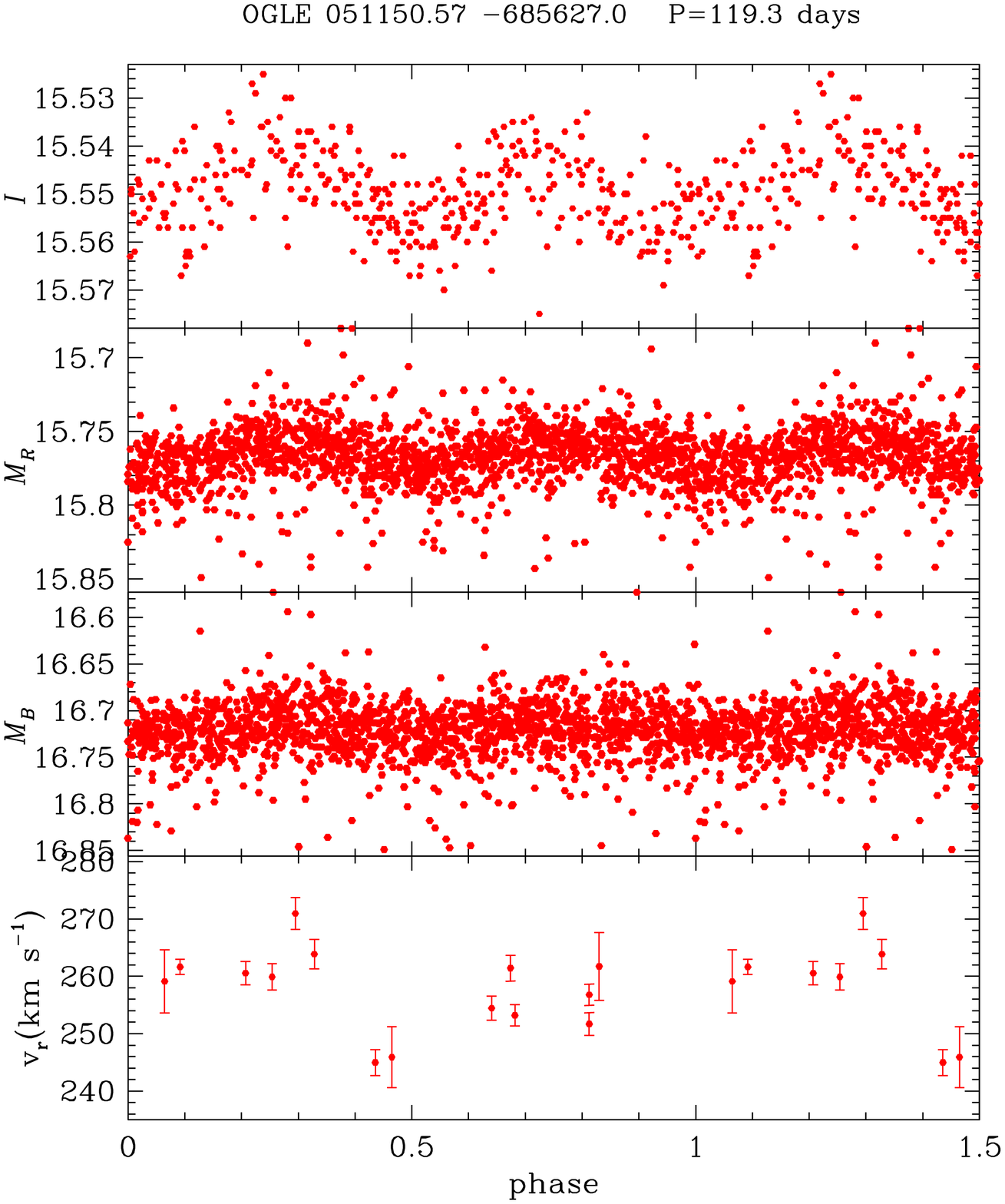}
\includegraphics[angle=0,width=0.34\textwidth,height=0.35\hsize]{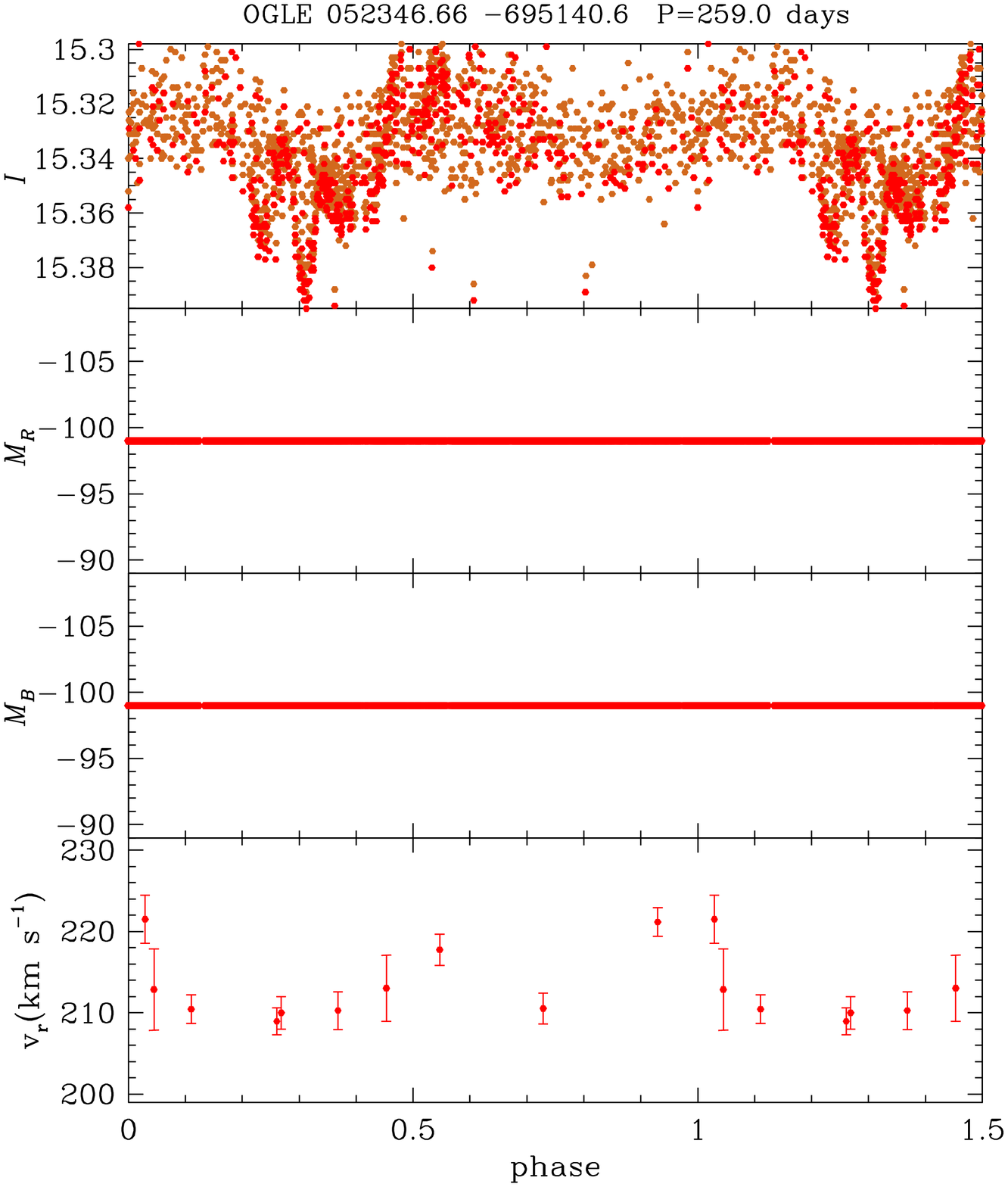}
\includegraphics[angle=0,width=0.34\textwidth,height=0.35\hsize]{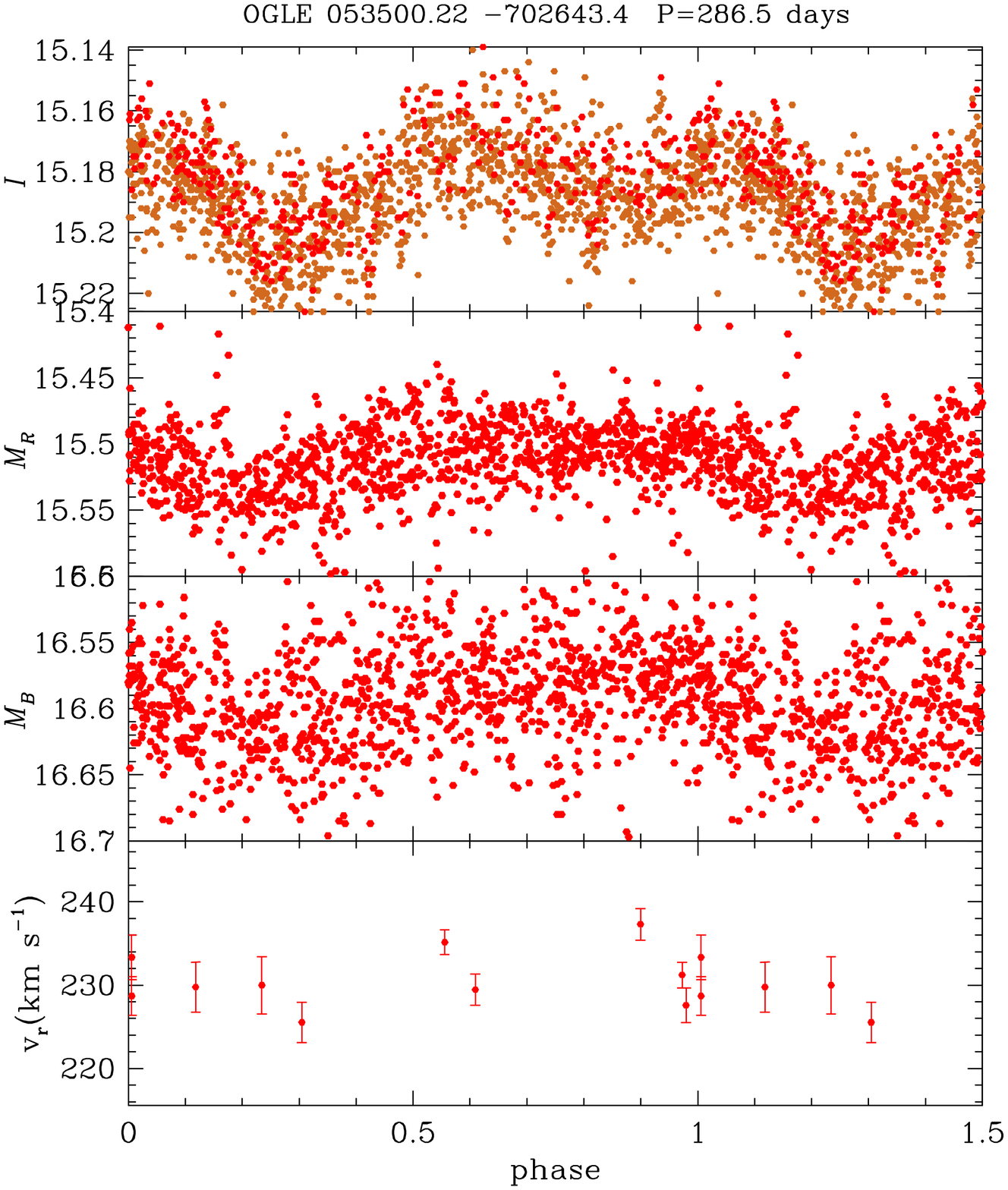}
\includegraphics[angle=0,width=0.34\textwidth,height=0.35\hsize]{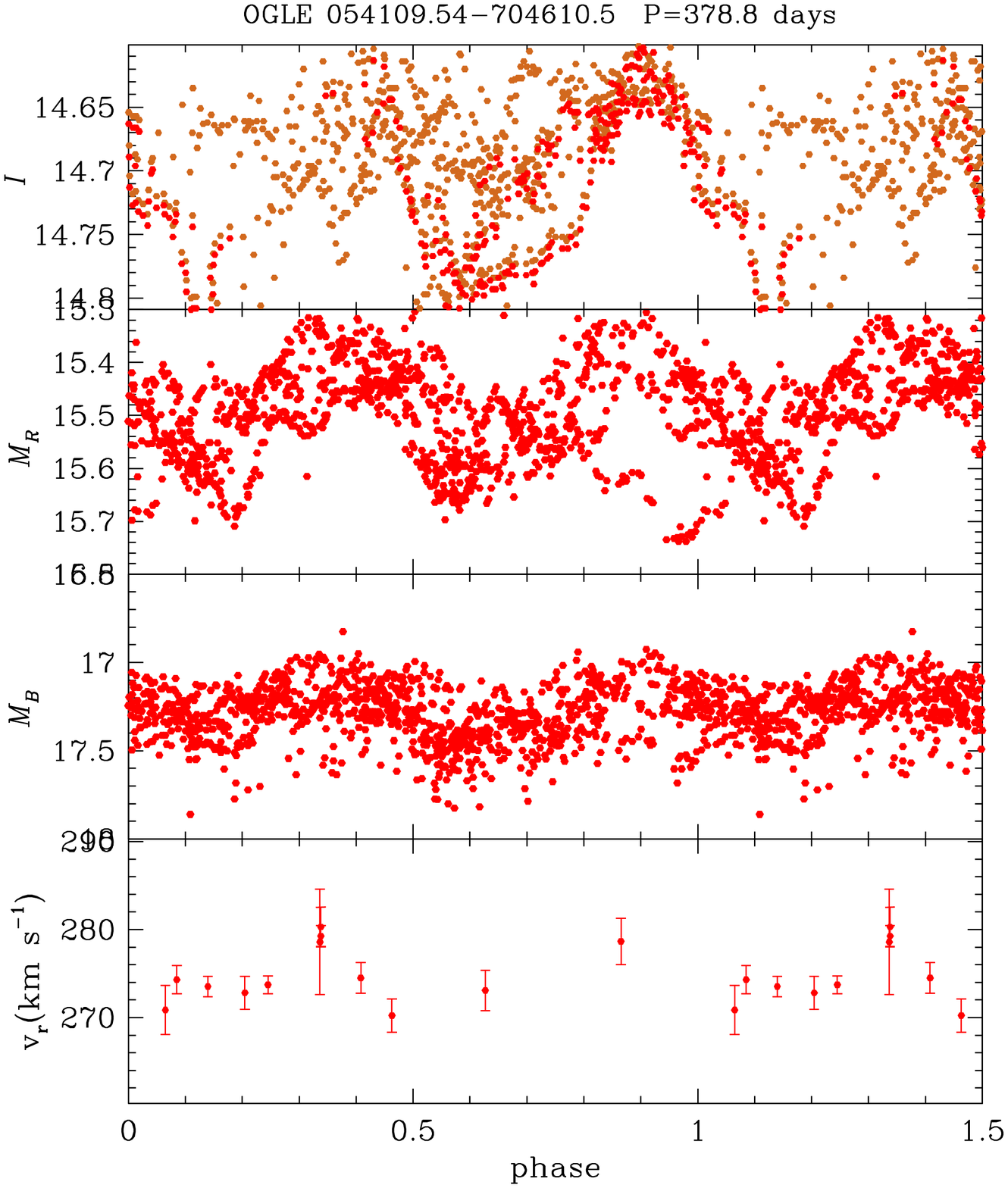}
\caption{Light and radial velocity curves of the 6 objects in the Appendix.
\label{fig_unsure_type}}
\end{figure*}

Here we present data for the 6 objects that we removed from the 86 
ellipsoidal candidates due to their poor velocity data relative to 
the noise. Properties of these 6 stars are given in Table \ref{sample-appendix}, 
and their light and velocity curves are presented in Figure \ref{fig_unsure_type}.
Their velocity data is given in Table~\ref{two_rv}.

\tabcolsep=0.2cm
\begin{deluxetable*}{lllllllllccccl}
\tablecaption{Properties of Objects Rejected\label{sample-appendix}}
\tablewidth{0pt}
\tablehead{
\colhead{No}&      \colhead{Object}&      \colhead{$P$}&
\colhead{$I$}&     \colhead{$V$}&         \colhead{$M_B$}&
\colhead{$M_R$}&   \colhead{$K$}&       \colhead{$\Delta{I}$}& \colhead{$\Delta{\rm RV}$}&
\colhead{$L$}&     \colhead{\teff}&       
\colhead{$R$}&     \colhead{Remark}\\
\colhead{}&        \colhead{(OGLE II Name)}& \colhead{(day)}&
\colhead{(mag)}&   \colhead{(mag)}&          \colhead{(mag)}&
\colhead{(mag)}&   \colhead{(mag)}&          \colhead{(mag)}&  \colhead{(\kms)}&
\colhead{(\lsolar)}&  \colhead{(K)}&       
\colhead{(\rsolar)}&  \colhead{} }
\startdata
1&050716.07-693259.7&  563.3& 14.81& 16.18&      -&      -& 13.09 & 0.025 & -  & 1571  &  4368 &    70&  1\\
2&051111.34-693714.9&  147.7& 15.69& 17.41&      -&      -& 13.59 & 0.060 & -  &  772  &  3984 &    59&  1\\
3&051150.57-685627.0&  119.3& 15.55& 16.91&  16.73&  15.77& 13.80 & 0.025 & -  &  804  &  4344 &    51&  1\\
4&052346.66-695140.6&  259.0& 15.34& 16.77&      -&      -& 13.54 & 0.050 & -  &  983  &  4290 &    57&  1,+sr,ogle3\\
5&053500.22-702643.4&  286.5& 15.19& 16.71&  16.60&  15.52& 13.34 & 0.060 & -  & 1141  &  4225 &    64&  1,ogle3\\
6&054109.54-704610.5&  378.8& 14.81& 17.17&  17.30&  15.55& 12.05 & 0.200 & -  & 2298  &  3517 &   130&  1,ogle3\\
\enddata
\end{deluxetable*}

\begin{thebibliography}{}
\bibitem[Allen(1984)]{1984PASAu...5..369A} Allen, D.~A.\ 1984, Proceedings of the Astronomical Society of Australia, 5, 369 
\bibitem[Adams, Wood, \& Cioni(2006)]{2006MmSAI..77..537A} Adams E., Wood P.~R., Cioni M.-R., 2006, MmSAI, 77, 537 
\bibitem[Belczy{\'n}ski et al.(2000)]{2000A&AS..146..407B} Belczy{\'n}ski, K., Miko{\l}ajewska, J., Munari, U., Ivison, R.~J., \& Friedjung, M.\ 2000, \aaps, 146, 407 
\bibitem[Cutri et al.(2003)]{2003yCat.2246....0C} Cutri R.~M., et al., 2003, yCat, 2246, 0 
\bibitem[Dopita et al.(2007)]{2007Ap&SS.310..255D} Dopita M., Hart J., McGregor P., Oates P., Bloxham G., Jones D., 2007, \apss, 310, 255 
\bibitem[Dopita et al.(2010)]{2010Ap&SS.327..245D} Dopita M., et al., 2010, \apss, 327, 245 
\bibitem[Duquennoy \& Mayor(1991)]{1991A&A...248..485D} Duquennoy A., Mayor M., 1991, \aap, 248, 485
\bibitem[Fraser, Hawley, \& Cook(2008)]{2008AJ....136.1242F} Fraser O.~J., Hawley S.~L., Cook K.~H., 2008, \aj, 136, 1242 
% \bibitem[Hall(1990)]{1990AJ....100..554H} Hall D.~S., 1990, \aj, 100, 554 
\bibitem[Houdashelt et al.(2000a)]{2000AJ....119.1424H} Houdashelt M.~L., Bell R.~A., Sweigart A.~V., Wing R.~F., 2000a, \aj, 119, 1424
\bibitem[Houdashelt et al.(2000b)]{2000AJ....119.1448H} Houdashelt M.~L., Bell R.~A., Sweigart A.~V., 2000b, \aj, 119, 1448
\bibitem[Ita et al.(2004)]{2004MNRAS.353..705I} Ita Y., et al., 2004, \mnras, 353, 705 
\bibitem[Kenyon(1986)]{1986syst.book.....K} Kenyon, S.~J.\ 1986, Cambridge and New York, Cambridge University Press, 1986, 295 p.  
\bibitem[Keller \& Wood(2006)]{2006ApJ...642..834K} Keller S.~C., Wood P.~R., 2006, \apj, 642, 834
\bibitem[Nicholls et al.(2009)]{2009MNRAS.399.2063N} Nicholls C.~P., Wood P.~R., Cioni M.-R.~L., Soszy{\'n}ski I., 2009, \mnras, 399, 2063 
\bibitem[Nicholls et al.(2010)]{2010MNRAS.405.1770N} Nicholls C.~P., Wood P.~R., Cioni M.-R.~L., 2010, \mnras, 405, 1770
\bibitem[Nicholls \& Wood(2012)]{2012MNRAS.421.2616N} Nicholls C.~P., Wood P.~R., 2012, \mnras, 421, 2616 
\bibitem[Nie, Wood, \& Nicholls(2012)]{2012MNRAS.423.2764N} Nie J.~D., Wood P.~R., Nicholls C.~P., 2012, \mnras, 423, 2764 
\bibitem[Mikolajewska et al.(1997)]{1997A&A...327..191M} Mikolajewska, J., Acker, A., \& Stenholm, B.\ 1997, \aap, 327, 191 
\bibitem[Madore(1982)]{1982ApJ...253..575M} Madore B.~F., 1982, \apj, 253, 575 
\bibitem[Raghavan et al.(2010)]{2010ApJS..190....1R} Raghavan D., et al., 2010, \apjs, 190, 1
\bibitem[Rieke \& Lebofsky(1985)]{1985ApJ...288..618R} Rieke G.~H., Lebofsky M.~J., 1985, \apj, 288, 618
\bibitem[Schlegel, Finkbeiner, \& Davis(1998)]{1998ApJ...500..525S} Schlegel D.~J., Finkbeiner D.~P., Davis M., 1998, \apj, 500, 525
\bibitem[Stellingwerf(1978)]{1978ApJ...224..953S} Stellingwerf R.~F., 1978, \apj, 224, 953 
\bibitem[Soszy\'nski et al.(2004)]{2004AcA....54..347S} Soszy\'nski I., et al., 2004, AcA, 54, 347
\bibitem[Soszynski et al.(2007)]{2007AcA....57..201S} Soszynski I., et al., 2007, AcA, 57, 201
\bibitem[Wood et al.(1999)]{1999IAUS..191..151W} Wood P.~R., et al., 1999, IAUS, 191, 151
\bibitem[Wilson \& Devinney(1971)]{1971ApJ...166..605W} Wilson R.~E., Devinney E.~J., 1971, \apj, 166, 605 
\bibitem[Wilson et al.(2009)]{2009ApJ...702..403W} Wilson R.~E., Chochol D., Kom{\v z}{\'{\i}}k R., Van Hamme W., Pribulla T., Volkov I., 2009, \apj, 702, 403
\bibitem[Wilson(1979)]{1979ApJ...234.1054W} Wilson R.~E., 1979, \apj, 234, 1054 
\bibitem[Wilson(1990)]{1990ApJ...356..613W} Wilson R.~E., 1990, \apj, 356, 613 
	
\end{thebibliography}
\end{document}